\documentclass[twocolumn,showpacs,preprintnumbers,amsmath,amssymb,amsmath]{revtex4-1}

\usepackage{graphicx}
\usepackage{dcolumn}
\usepackage{bm}
\usepackage{hyperref}
\usepackage{epsf}
\usepackage{float} 
\newcommand{\eq}[1]{(\ref{#1})}
\newcommand{\la}{\label}
\newcommand{\bea}{\begin{eqnarray}}
\newcommand{\eea}{\end{eqnarray}}
\newcommand{\beq}{\begin{equation}}
\newcommand{\eeq}{\end{equation}}
\newcommand{\be}{\begin{equation}}
\newcommand{\ee}{\end{equation}}
\newcommand{\ii}{{\rm{i}}}

\newcommand{\vv}{{\rm v}}
\newcommand{\uu}{{\rm u}}
\newcommand{\pp}{{\rm p}}
\newcommand{\p}{\partial}

\def\XXint#1#2#3{{\setbox0=\hbox{$#1{#2#3}{\int}$ }
\vcenter{\hbox{$#2#3$ }}\kern-.5\wd0}}

\usepackage{amsmath}
\usepackage{amsfonts}
\usepackage{varioref}

\begin{document}

\title{  Hydrodynamics of Euler incompressible fluid and the  Fractional Quantum  Hall Effect}

\author{P. B. Wiegmann }
 \affiliation{ Department of Physics, University of Chicago, 929 57th St, Chicago, IL 60637}
\date{\today}

%

\date{\today}

\begin{abstract}

We show that  the Fractional Quantum Hall Effect  can be phenomenologically  described as a special flow of    a quantum   incompressible  Euler liquid. This flow consists of a large number of vortices of the same chirality. In this approach each vortex is identified with an electron while the fluid is neutral. We show that the Laughlin wave function naturally emerges as a stationary flow  of the system of  vortices in quantum fluid dynamics. Subtle features of  FQHE such as   effects  of  Lorentz shear stress, the spectral function,  the   Hall current
in a modulated  landscape, etc., naturally follow from the hydrodynamics approach. In the paper we develop the hydrodynamics of the vortex liquid, and able  consistently quantize it. As a demonstration of the efficiency of the hydrodynamics we briefly  discuss some new results \end{abstract}

\maketitle

\noindent In  the Fractional Quantum Hall  regime (FQH)  electrons form an perplexing quantum liquid. Some major characteristics of this liquid are well established  theoretically and experimentally: the liquid is incompressible \cite{L},  almost  dissipation-free \cite{T,T1},  the Hall conductance  is  quantized 
\cite{T1}, excitations   are vortices which carry  a fraction of the negative electronic charge \cite{L}, neutral bulk excitations  are   gapped \cite{T}. More subtle properties in focus of recent interest is  the Lorentz shear force, or odd (aka anomalous viscosity or Hall viscosity)
\cite{Avron,ReadHV,Tokatly,Son,Haldei, Abanov}        

A natural approach to   FQHE advocated in a seminal paper \cite{GMP},  is quantum hydrodynamics.   Quantum hydrodynamics is based on a set of fundamentally restrictive assumptions that long and slow waves are described exclusively by a closed system  of conservation laws. Hydrodynamics often  difficult to derive from {\it\ ab initio} microscopic basis, but once developed has a predictive power and could be tested against the known properties.  

Here we extend this approach. First, irrespectively
from the FQHE, we develop the  quantum hydrodynamics of the vortex flow in the  two-dimensional  incompressible Euler  fluid. Then we  see how  major concepts of the FQHE, such as Laughlin's wave function and fractionalization of the Hall conductance and excitations  emerge in  the Euler hydrodynamics. Then we  obtain more   subtle properties of FQHE, the Lorentz
shear force and  anomalous viscosity. All  naturally  follow from the hydrodynamics of the quantum Euler fluid.

This is, of course,  not  a accidental coincidence,  but rather  a confirmation of the  conceptional viewpoint that the major   properties of the FQHE are governed by    symmetries and the underlying geometry of the  states. The hydrodynamics  reveals and clarifies the symmetries. 

As a  demonstration of  the effectiveness of the hydrodynamic approach we compute, the spectral function and  linear response of the electronic fluid to non-uniform electric and magnetic fields, the density profile at the quasi-hole and  accumulation of charge on a curved surface. Some of these results are new. 

We consider only   Laughlin's cases, where  fraction \(\nu^{}\)
is
an inverse of an odd integer, say 1/3. Extensions of the
hydrodynamic approach  to FQH states, possessing
 external symmetries will be discussed elsewhere.

 Historically the quantum hydrodynamics   goes back to studies the superfluid
helium by Landau
\cite{Landau} and
Feynman \cite{Feynman}.  A quest for the hydrodynamics of the FQH liquid has been   originated in \cite{GMP}.  Earlier approaches to FQHE  \cite{Kivelson,Dung,Read89,Stone} were essentially related to hydrodynamics as  explained in \cite{Stone}. Hydrodynamics of  FQH liquid  is  a focus of  a renewed
 interest \cite{Haldei,Avron,Tokatly,ReadHV,Son, Abanov}. 

Among vast variety of  flows in  the incompressible  Euler fluid only one special class of flows is relevant to FQHE. This is a \emph{turbulent flow} where vorticity is proportional to the volume. Such flow consists of a dense system of quantized vortices, all  oriented in one direction. We will be interested in a regime where  vortices    themselves constitutes a  liquid,  the \emph{vortex  liquid}.


In the paper  we  develop  the hydrodynamics
of \! such vortex fluid   in a  close analog of the Feynman theory of rotating  superfluid helium\! \cite{Feynman}, see also \cite{Khalat}, where a similar setting occurs in the regime when the lattice of vortices is melted. The difference, however is crucial: in contrast to helium, the FQH liquid is \emph{incompressible}.

 Then we observe that properties of the vortex liquid are identical to the  FQH electronic liquid.  In other words, external forces  applied  to  the vortex liquid (not to
the liquid itself, but  to vortices)  generate the same motion as FQH-electronic liquid under electric and magnetic  fields.  

This observation suggests   a phenomenological picture of FQHE:  collective electronic states are localized on vortices, the topological configurations, of a neutral  
incompressible  liquid. The liquid itself is a neutral agent which mediates
interaction between electrons. The similar picture is known in organic conductors (see e.g., \cite{K}). There electrons occupy the core of topological configurations (kinks) of ion displacements, the neutral field mediating electronic interaction.\

Quantization of incompressible hydrodynamics is  a subtle matter due to it
non-linear nature. In this paper we present  perhaps the first example of
consistently quantized hydrodynamics. We achieve it through quantization
of
Kirchhoff equations for   vortices dynamics. 
\smallskip

\noindent\ We start by a brief  discuss the energy and length scales in the fluid mechanics and FQHE.

 In hydrodynamics only   few basic principles, symmetries and
  phenomenological parameters  suffice to formulate   fundamental
equations. The  phenomenological parameters of the quantum  hydrodynamics  is the circulation of each vortex \(2\pi \Gamma\). The characteristic of the flow is   the mean density of vortices \(\bar\rho\). We  assume that the liquid performs a solid rotation with  the frequency \(\Omega=\pi\Gamma\bar\rho\),   such that the net vorticity vanishes.   The energy  of the solid rotation \(\hbar\Omega\) is  the only energy scale of the flow.
 
On the other hand  the only energy scale in the FQHE is given by the gap in the excitation spectrum \(\Delta_\nu\), typically \(\Delta_\nu\sim 10K\) \cite{T}. This scale is   controlled by  the Coulomb interaction. It is customary to introduce a scale of mass associated with the this energy setting \(m_\nu\sim\hbar^2/\Delta_\nu\ell^2)\), where  \(\ell=\!\sqrt{\!\hbar
/\!e B}\!\) is magnetic length. The very existence of the FQH state requires
that the gap to be less than the cyclotron frequency \(\Delta_\nu\ll\hbar\omega_c
 \), so that all  states are confined on the lowest Landau level. This means that \(m_\nu \)
 exceeds the band electronic mass \(m_b\). 

In the absence of  other scales  it appears that   \(\hbar\Omega\) and \(\Delta_\nu\), and that \(\bar\rho\) and \(\ell^{-2}\) are of the same order.   Then the scale of the vortex circulation  is \( \Gamma  \sim\hbar/m_\nu\) and \(m_\nu\) is the inertia of the fluid.

The states  on the lowest Landau level are 
holomorphic. We will see  that  this property  means that the electronic  liquid is incompressible.  Velocity is divergence-free.   

Existence of the energy scale within the Landau level is the physical
input justifying the hydrodynamics of the FQHE. For that reason the hydrodynamics description  does not extend to the integer  case, where interaction is weak and  the cyclotron energy is the only scale. The role of interaction could be   seen within the hydrodynamics itself.
An incompressible liquid does not possess  linear  waves except on the edge \cite{LV}. All  flows are  non-linear.

 Fractional
and
Integer Hall effects can be treated in parallel and  within the hydrodynamic approach only in the \emph{topological
sector} singled out by the limit \(\Delta_\nu\to\infty\).  Flows in this sector are steady, such as the Hall current. 

After these comments we turn to the Euler hydrodynamics. 
We start from the classical case. \smallskip

\noindent 
Incompressible   \(\nabla\cdot{  u}=0\) flows in two dimensions 
 are fully characterized by its vorticity \(\omega=\nabla\times u\), where \(u\) is the fluid velocity. Vorticity obeys a single  equation, which in the case of inviscid  fluid has a simple geometrical meaning: the material derivative of the
vorticity vanishes  
\begin{align}\la{E}{D_t}\omega\equiv\left(\frac{\partial}{\p t} +u\cdot \nabla\right)\omega=0,\quad   \nabla\cdot{  u}=0. 
\end{align}Vorticity is  transported
 along     divergence-free velocity. 

In the class  of Helmholtz solutions the complex velocity   is a meromorphic function. In the rotating frame  
 \begin{align}\label{k}
\uu(z,t)=-\ii\Omega \bar z+\ii\sum^N_{j=1 }\frac{\Gamma_j}{z-z_j(t)} .
\end{align}
Here  \(\Gamma_j\) and   \(z_j(t) \) are circulations and positions
of vortices. The Kelvin theorem insures that the number of vortices \(N\) and their circulations \(\Gamma_i\) do not evolve.  

A substitution of the "pole Ansatz" into the Helmholtz equation \eq{E}
expresses the velocity
of vortices 
as a sum the Magnus forces
exerted by other vortices   \begin{align}\label{u}
\vv_i\equiv  \dot {\overline z}_i=\ii\Omega\bar z_i+\ii\sum^N_{i\neq j }\frac{\Gamma_j}{z_i(t)-z_j(t)}.
\end{align}
This  dynamical system is called  Kirchhoff equations  \cite{Newton}.  It   replaces the non-linear PDE \eqref{E}.  Equations describe chaotic motions  if \(N>3\). In a proper limit of large \(N\)  and small \(\Gamma\) solutions   approximate virtually any flow. 

  \smallskip
 
\noindent  We will be interested in the system of large number of vortices \(N\to\infty\)  , the turbulent flow, and specifically in the {\it chiral flow}, where   all vortices have the same (minimal) circulation
 \(\Gamma_i=\Gamma\).

In this limit the vortex system must be treated as a liquid itself.    

 In the turbulent flow  we distinguish two types of motion: a fast motion of the fluid around vortex cores, and a slow motion of vortices fluid. In particular, in  the ground state of   the vortex liquid, the vortices  do not move, but the fluid does. In the 
  stationary flow   vortices are distributed uniformly
 with the mean density  \(\bar\rho=\/\Omega/(\pi
\Gamma)     \). 

Kirchhoff equations are scale invariant. They do not change under a dilatation \(z_i\to\lambda z_i,  t\to\lambda^2 t,\bar\rho\to\lambda^{-2}\bar\rho\) and for that reason do not consist of any  energy scale.   In order to write the Hamiltonian one needs to introduce an {\it\ ad hoc} scale of energy. Bearing in mind application to FQHE, we set it  to be   \(\Delta_\nu\). Then the Hamiltonian   \begin{align}\mathcal{H}=\Delta_\nu\sum_i\left(\pi\bar\rho|z_i|^2-\sum_{j\neq i}\log|z_i-z_j|^2\right)\la{H1}\end{align}  and the Poisson brackets 
  \(\{\bar z_i,\,z_j\}_{P.B.}=\frac{\Gamma}{\ii\Delta_\nu}\delta_{ij} \) reveal the Kirchhoff equation \eq{u}. The scale \(\Delta_\nu\) disappears from the equations.    
\smallskip
 \noindent

Now we proceed with the quantization. The first step is to replace the Poisson brackets   by the commutators \begin{align}\label{C}
\{\bar z_i,\,z_j\}_{P.B.}\rightarrow[\bar z_i,\,z_j]=2\ell^2\delta_{ij},
\end{align}
where we denote  \(2\ell^{2}=\hbar\Gamma/\Delta_\nu\). It has a dimension of area. The ratio between this scale and the area per particle \(\nu=2\pi\bar\rho\ell^2\) is the  dimensionless
  semiclassical parameter. We will see in  a moment that \(\nu\) appears to be the filling fraction, and  \(\ell\) to be the magnetic length.

At the next step we must specify the space  of states. We assume that states are  holomorphic polynomials of \(z_i\). Then operators
\(\bar z_i\)  are  canonical momenta \be\la{r}\bar  z_i  =2\ell^2\p_{z_i}.\ee The last step is to specify the inner product. We impose  the \emph{chiral condition}: operators \(\bar z_i\)  and \(z_i\) are assumed to be Hermitian conjugated
\begin{align}\la{hermitte}
\text{\small chiral condition}:\quad &\bar z_i^\dag=z_i.&
\end{align} The conditions   \eq{r} and \eq{hermitte} identify the set of  states with the Bargmann space \cite{GMP,Bargmann}:\ the Hilbert space  of analytic polynomials \(\psi(z_1,\dots,z_N)   \) with the inner product 
\begin{align}\la{B}
\langle\psi'|\psi\rangle=\int e^{-\sum_i\frac{|z_i|^2}{2\ell^2}}
\overline{\psi^\prime}\psi  \,d^2z_1\dots d^2z_N  \end{align}
Eqs. (\ref{u},\ref{r}) help to write quantum velocity  operators as\begin{align}\la{v}
   \pp_i=-\ii\hbar({ \p}_{z_i}-\sum_{j\neq
i}\frac{\beta}{z_i-z_j}),\quad\beta=\nu^{-1},
\end{align}
 where we set \(\pp_i=m_\nu\vv_i\) and the effective mass  \(m_\nu=\hbar/(\nu\Gamma) \). Operators \(\pp_i\) are the many-body version of the guidance center coordinates or coordinates of magnetic translations.  
\smallskip

\noindent 
A the stage the    Kirchhoff equations are readily identified  with the  FQHE in a disk geometry. There   the electronic droplet occupies a volume
confined by a weak potential.
\smallskip

We recall  that
 the Bargmann space  is  just another way to say  that all states belong to the lowest Landau level.
In that representation the wave functions are written in the radial gauge with respect to a marked point (the origin) inside  the droplet (see e.g., \cite{GMP} for details). Apart from the factor \(\exp (-\sum_i|z_i|^2/{2\ell^2})
\), treated as a measure, the states are holomorphic polynomials.

Let us determine the ground state of the vortex liquid. There all velocity vanish   \(\pp_i\psi_0=0\).   The common solution of  the  set of  first order PDEs is the  Laughlin function  \begin{align}\la{L} \pp_i\psi_0=0,\quad\psi_0=\prod_{i>j}(z_i-z_j)^\beta,\quad\quad\nu=\beta^{-1}.\end{align}   
The wave function is single valued if \(\beta\) is integer. Depending whether $\beta$ is chosen to be odd or even  integer   the vortices are Fermions or Bosons. In particular, $\beta=2$ is believed to describe the rotating  Bose  condensate of trapped  atoms. At \(\beta=3\) we obtain the Laughlin \(\nu=1/3\) state.

We observe that   the Laughlin states, fermionic or bosonic alike, naturally  emerge from the quantum   hydrodynamics of the vortex fluid.
In this approach, the fraction appears as a parameter of the quantization.

In the hydrodynamics  interpretations  "particles"  entered into the Laughlin function are vortices of the incompressible  fluid.
In the FQHE  particles are electrons, with electric charge. To complete the hydrodynamics description we must identify electric and magnetic field  as field acting on vortex cores. 
To this end we add a potential  \(\sum_i U(r_i)
\)  to the energy \eq{H1}, where \(r_i\) are coordinates
of vortices. It exerts the  force  \(-\ii[U,\bar z_i]= \ii2\ell^2\p_{z_i}
U \) added to  the Kirchhoff  equations\begin{align}\la{15}
\vv_i=-\ii\Omega\bar z_i+ \ii \sum_{j\neq
i}\frac{\Gamma}{z_i-z_j}+\ii\ell^2 e\mathrm{E},
\end{align}
where \(eE=-\nabla U\) plays a role of  the electric field.
The electric field acts normal to velocity. It does not  accelerate the flow since  vortices have no  "mass". It must not be confused with the \(m_\nu=\hbar/(\nu\Gamma)\), the inertia of the fluid. 
Thus we  identify the angular velocity with the cyclotron frequency
of vortices \(\Omega\!=\!eB\!/\!m_\nu=(m_b/m_\nu)\omega_c\ll\omega_c\),\(\ell\) with the  magnetic length  \(\ell=\!\sqrt{\!\hbar
/\!e B}\!\), and  \(\nu\!=\!\hbar/(m_\nu\Gamma)  \) the filling fraction.

To illustrate the assignment electric charges  to vortices we invoke a similar  phenomena known in organic conductors  \cite{K}. There  electronic states are   localized on cores of  kinks  of ion displacements and move together if the motion is adiabatic. The kinks are the topological configurations of the 1D  phonon field. Here, in a very similar manner electronic states  are trapped by  vortices, the topological configurations in 2D. This is only the illustration. It does not explain a microscopical mechanism of attachment of the vortex circulation to the electron, but rather provides a hydrodynamics  interpretation to the commonly used concept of  the "flux attachment". 

Quantization of the Hall conductance elegantly follows from the Kirchhoff equations \eq{15}. Let us assume that the electric field is uniform and sum up all the equations. We obtain the relation between the e.m. current
and the electric field \(N^{-1}e\sum_i(\vv_i+ \ii \Omega \bar z_i)=\ii\ell^2 e^2\mathrm{E}\)  with the fractionally quantized conductance 
\(\sigma_{xy}^{}=\nu (e^2/h)\).
If  the electric field is not uniform, the Hall conductance possesses universal corrections described below.

The fractionalization of quasi-holes is another easy consequence of the  Kirchhoff equations. The quasi-hole \cite{L}
 is a state with the wave function \(\psi_h=\prod_i(z-z_i)\psi_0.\) 
 The operator \eq{v} acting on this state   is $$\pp_i\psi_h=-\ii\hbar({ \p}_{z_i}+ \frac{1}{z_i-z}-\sum_{j\neq
i}\frac{\beta}{z_i-z_j})\psi_0.$$
It shows that the Magnus force exerted by  vortices to the quasi-hole is
the     the  fraction \(\nu\) of the  forces between vortices and acts in the opposite direction.  Thus
 in the hydrodynamic interpretation   the quasi-hole appears  is a vortex
 with a fractional negative circulation \(-\nu\),  an anti-vortex, or a hole
 in the uniform
"Fermi sea" of vortices.

These arguments seem to justify (\ref{H1},\ref{C},\ref{B},\ref{15}) as complete minimal set of FQHE dynamics.


\smallskip

\noindent\ Our next goal is to obtain the  hydrodynamic description of the vortex fluid. From the hydrodynamics standpoint, the coordinates of vortices are treated as Lagrangian specification of  fluid parcels. To pass  to the Eulerian specification we  must consider   the macroscopic
conserved fields: the vortex density and the vortex flux    \begin{align}\la{45}
 &\rho(r)=\sum_{i}\delta(r-r_i)=\bar\rho+\frac{1}{2\pi\Gamma}(\nabla\times
u),\\&\mathcal{J}(r)=\sum_i\delta(r-r_i)\vv_ i,\la{455}
 \end{align}compute them and determine velocity through the relation   $$\mathcal{J}(r)=\rho(r)\vv(r)
.$$By construction the flux annihilates the 
the ground  state 
 $$\mathcal{J}|0\rangle=\langle 0|\mathcal{J^\dag}=0.$$

To the best of our knowledge this program has never been set up even for the classical fluids. Below we outline the major step.
To simplify the formulas we compute the flux classically. The quantum    result  is  the same, providing the ordering of operators is kept.

We  write the vortex  flux  
\begin{align}\la{152}\mathcal{J}=\sum_i\delta(r-r_i)[-\ii\Omega\bar  z_i+ \sum_{j\neq
i}\frac{\Gamma}{z_i-z_j}].\end{align}
and use the \(\bar\p\)-formula
\(\pi\delta=\bar\p(\frac 1z) \)  and the identity  $$2\sum_{i\neq
j}\frac{1}{z-z_i}\frac{1}{z_i-z_j}=(\sum_i\frac{1}{z-z_i})^2\!-\!\sum_i(\frac{1}{z-z_i})^2.$$
A simple computation yields the important relation between the vortex flux and  the vorticity flux  
  \begin{align}&\!\mathcal{J}=-\ii\rho\Omega\bar z+\!\ii\frac{\Gamma} {2}\bar\p[(\sum_i\frac{1}{z-z_i})^2\!-\sum_i\!\frac{1}{(z-z_i)^2}]\nonumber
\\&=\!\rho[-\ii\Omega\bar z+{\ii }\sum_j\frac{\Gamma}{z-z_j}]+\ii\frac {\Gamma} {2}\p\rho=\mathrm{\rho\uu}+\ii\frac
{\Gamma}{ 2} \p\rho\label{v1}. 
\end{align}
 The first term in  the r.h.s. is the vorticity flux \(\mathrm{\rho\uu}   \).  The second  is the anomalous term. It appears because the velocity of  the fluid \(u\) diverges at a core
of an isolated vortex
 (as  in (\ref{k})). However, velocities of  vortices are finite.
The  anomalous term   removes that singularity.

 In Cartesian coordinates the relation between velocity of the fluid and velocity of the vortex fluid reads (we denote 
\((\nabla\times)_a=\epsilon_{ab}\nabla_b)\)\cite{Calogerocomment}\begin{align}\la{shift}
   \mathcal{J}\equiv\rho v=\rho  u+\frac {\Gamma}{4} \nabla\times\rho=\rho u-\frac{1}{4\pi}\Delta u.
\end{align}
%

\noindent
%
The  meaning of the anomalous  term is  seen  from the geometric phase
of  the FQH
states.
That
is   the    phase
acquired by the state when a chosen  particle  moved around a closed  path encompassed all other particles.  In units of \(2\pi\) it equals to the number
 of zeros  of the wave function with respect
to  a chosen particle \(N_\phi\)  and equals to the number of fluxes of magnetic field \(N_\phi=(N-1)\beta\) in the disk geometry. 
The  "shift", i.e., the difference between \( N\) and \(\nu N_\phi\) is the contribution  of  the anomalous term. It can be seen as a result of integration of the shift relation \eq{shift} over a contour encompassed the droplet.
The condition \eq{shift}  is
the local version of  the "shift", the global relation between the magnetic flux and  the number of particles (see, e.g., \cite{shift2}). 

We see that the vortex flow  is  incompressible
  like the fluid itself and that the Helmholtz  equation
 \eq{E} emerges as the continuity equation for the vortex
liquid\begin{align}
\mathcal{D}_t\rho=0,\quad \mathcal{D}_t=\p_t+v\cdot\nabla, \quad  \nabla\cdot{  v}=0.
\la{cont}\end{align}

The relation \eq{shift} has
 far reaching consequences.
One of them is the 
Lorentz shear stress.

The rotating fluid  parcel experiences the Coriolis   force \(\rho F=-m_\nu\Omega\times(\rho   u)\). This force also acts on vortices. To find its action  we express it through  the velocity of the vortex fluid. With  the help of  the shift formula \eq{shift}  neglecting higher  orders in gradients we obtain \begin{align}\rho F\approx eB\times (\rho v)-\frac{\hbar}{4\nu}\bar\rho\nabla(\nabla\times v).\la{19} \end{align}The first term here is the familiar  Lorentz force, the second is the Lorentz shear force. The universal coefficient translated to this formula from \eq{shift} is the anomalous viscosity (aka  odd viscosity or Hall viscosity). 

The anomalous force  could be written as a divergence of the symmetric Lorentz shear stress tensor \(F_a=\nabla_b\sigma'_{ab}\).
which  is best written in terms of the stream
function
\begin{align}\sigma_{ab}'=\frac{\hbar}{2\nu}(\nabla_a\nabla_b-\frac
12\delta_{ab}\Delta)\Psi, \quad v=-\nabla\times\Psi.\end{align}    
The anomalous  stress is  conservative    and traceless. To  compare, the dissipative shear viscous tensor is given by the same formula where the stream function is replaced by the hydrodynamic potential.

Initially  introduced for the integer QHE in  \cite{Avron} it has been  extended
to  the  FQHE in   \cite{Tokatly,ReadHV}. In fact, the Lorentz shear force is 
 the hydrodynamic and also  classical phenomena reflecting the discreteness of vortices.

The anomalous force could be visualized 
as  a strain of  orbits of the fluid around the vortex cores by the shear flow. The flow elongate them
normal to the shear squeezing together  
 flow lines with different velocity  exerting additional force toward the boundary.\smallskip

  We see that   the Lorentz shear force naturally emerges in the hydrodynamics of the vortex flow. To obtain further applications, we need the hydrodynamic form  {\it chiral condition} \eq{hermitte}. From now on  we set \(m_\nu=1\), or \( \Gamma=\hbar/\nu\).

\smallskip

\noindent

In classical incompressible fluids  the position of vortices determine their velocities, as it is seen from  the classical Kirchhoff
equation. The chiral condition \eq{hermitte} insures that the same is true in the quantum case.  In hydrodynamics terms this means that  the vortex flux \(\mathcal{J}\)  and the velocity are determined by the density of vortices
 \(\rho\). This is the chiral  consistency condition we want to obtain.  It reflects the holomorphic nature of states, or equivalently  the incompressibility of the fluid, or  that all states belong to the first Landau level.

The chiral consistency relation is obtained when we apply "normal ordering" to the shift equation  \eq{v1}. This means to place the   holomorphic operator of velocity \(\uu\)  to the left next to the "bra"- anti-holomorphic state. Then \(\uu \) possesses no differential operators and acts  classically as a solution of \eq{45}
$$
\langle...|\uu=\langle...|(-\ii h/\nu)^{}\p\varphi, \quad\Delta\varphi=-4\pi(\!\rho-\bar\rho).
$$
 The normal ordering is achieved with the help  of canonical equal point commutation relation \([\uu(r),\rho(r)]=\ii\hbar\p\rho(r)\). It \(\)essentially changes the coefficient in the shift equation \eq{shift}
 \begin{align}
\mathcal{J}=\uu\rho-\ii
\hbar\p\rho=\ii\frac{h}{\nu}\rho\left(\p\varphi+(\frac{1}{2}-
\nu)\p\rho\right)\la{20}
\end{align}  This is the {\it\ chiral constituency condition} \cite{WZ}. It expresses the flux in terms of one and  two-point density functions. The consistency condition especially efficient in the topological sector, where physics is bound to the leading gradients. In this regime we may treat the relation \eq{20} classically as we assume below. 
In the remaining part of the letter we list an incomplete set of applications
emphasizing the role of the anomalous term.

\paragraph{Flux attachment and the profile of the quasi-hole.} Let us divide (\ref{19},\ref{20}) by \(\rho\)  and take a curl of \eq{20} \begin{align}
 &\la{211}
\nabla\times v=\!\frac{h}{\nu}[\rho-\bar\rho+\frac{ 1}{4\pi}(\frac{1}{2}-
\nu)\Delta\log\rho],
\\&
F\approx eB\times v+\frac{\hbar}{2\nu}(\frac{1}{2}-
\nu)\nabla(\nabla\times v).\la{212}
\end{align} 
Would the last term in \eq{211} 
be ignored the vorticity of the flow  follows the density of particles times the filling fraction. This condition has been suggested  in \cite{Stone} as a basis for the hydrodynamics of FQHE and reflect to a popular picture that FQH states are electronic states
with  attached additional magnetic flux. The  {\it anomalous
term} corrects this concept. 

In the linear approximation a modulation of the density    \(\rho_k
 =\sum_i e^{\ii k\cdot r_i}\) causes  velocity
  \begin{align}\la{26}\mathcal{\vv}_{k\!}\approx\frac
h\nu \frac{\rm 
k}{ k^2}\left(1\!-\!\frac{1}{2\nu}(\frac{1}{2}-\nu)(k\ell)^2\right
)\rho_k.\end{align}
Eq. (\ref{211})  can be used to find  density  profiles for various coherent states. For example a quasi-hole  is  a source   \(\nabla\times v=-h\delta(r-r_0)\) in the  equation \eq{211}. Outside the core and in the leading gradients the quasi-holes causes a modulation \cite{vortex}.  $$\rho_k^{(h)}\approx(\bar\rho-\nu)\delta_{k,0}- \left(\nu-\!\frac{1}{2}(\frac{1}{2}-\nu)(k\ell)^2\right
).$$
\smallskip 
\paragraph{Structure function.}
The structure function \(s_\nu(k)=N^{-1}\langle0|\rho_k\rho_{-k}|0\rangle\) is the correlation of    density modes. To compute it we use the hydrodynamics commutation relation
\([\mathrm{\mathcal{\vv}}(r),\rho(r')]=-\ii\hbar\p\delta_{rr'}\) followed from  
\eq{455} and \eq{v}.  We recall   that the holomorphic velocity annihilates the "ket" vacuum.  Therefore \(\langle 0|\vv_k,\rho_{-k'}|0\rangle=\frac
12\hbar {\rm k}\delta_{k,k'}\). Substitute \eq{26} there and obtain the celebrated result of \cite{GMP} (see \cite{confusion})\begin{align}\la{80}
s_\nu(k)\approx \frac{1}{2}(k\ell)^2\left(1+\frac{1}{2\nu}(\frac{1}{2}-\nu)(k\ell)^2\right).
\end{align} 
\paragraph{Non-uniform electric field}   At the steady state the electric field balances Lorentz force plus the Lorentz shear force \eq{212} balanced \(F=eE\).  Solution of this equation
gives the Hall current \(e\bar\rho v_k=\sigma_{xy}(k) E_k\). The   Hall conductance acquires the universal correction \cite{Son}:
\begin{align}
 \sigma_{xy}(k)=\frac{\nu
e^2}{h}\left(1+ \frac{1}{2\nu}(\frac{1}{2}-\nu)(k\ell)^2\right).\la{30} 
\end{align}  \paragraph{Non-uniform magnetic field.}Similar relation  occurs between the density and a non-uniform magnetic field. A non-uniform magnetic field enters into the relation   \eq{211}
 through   the mean density   \(\bar\rho=\frac{\nu}{h
}eB\). At the ground state where velocity vanishes the   \eq{211}
becomes the Liouville-like equation for the density. In the leading   approximation in gradients we obtained a generalization of  the Streda formula  \(e\langle
0|\rho_k|0\rangle =\sigma_{xy}(k)B_k  \) for a weakly non-uniform magnetic field: \(\sigma_{xy}(k)\) is the
same as in \eq{30}.

\paragraph{Accumulation of charges in curved space.} \smallskip

Anomalous properties of FQHE are   seen in a curved space. Here we mention just one. In  a curved space the density (the number of particles per unit area \(\rho\sqrt g dzd\bar z) \) is not uniform but rather depends on the curvature \begin{align}
\rho=\bar\rho+\frac{1}{4\pi}R+\mathcal{O}(\ell^2\Delta R).\la{261}
\end{align}
 The first term of the gradient expansion in the curvature
follows from the shift
formulas \eq{shift}
In the curved space the density transformed as \(\rho\to \rho\sqrt g\). Under this transformation the anomalous term in \(\eq{shift}\) acquires an addition \(\frac{\hbar}{4\nu}\nabla\times\sqrt g\) which yields the term \(-\frac{1}{2\pi}\frac{1}{\sqrt g}\Delta\log\sqrt g\)  in the
r.h.s. of \eq{211} and subsequently \eq{261}.
Recall that \(R=-\frac{2}{\sqrt
g}\Delta\log\sqrt g\) is the Gaussian curvature. The next term in the expansion \eq{261} is also universal, but 
requires a more involved analysis.

 Particles/vortices accumulate at curved parts being pushed there by the Lorentz shear force. For  example, a cone   with the deficit angle \(\alpha\) possesses  extra \(\alpha/4\pi \) particles located right at the vertex. 

Eq.\eq{261} can be checked against the known formula for the number of particles at the Laughlin state on a Riemannian manifold. Integrating \eq{261} and using Gauss-Bonnet theorem we obtain   \(N=\nu N_\phi+\frac 12 \chi\) , where \(\chi\) is Euler characteristic.

Discussions of hydrodynamics of quantum liquids with A. G. Abanov, I  Rushkin, E. Bettelheim and T. Can  and their help are acknowledged. 
 The author
 thanks the International Institute of Physics (Brazil) and Weizmann Institute of Science (Israel) for the hospitality during the
completion of the paper. The work was supported in parts by NSF DMS-1206648, DMR-0820054, BSF-2010345 and John Templeton Foundation.
 

\begin{thebibliography}{10}
\bibitem{vortex}A    similar
 equation  but with  the term  \(-\frac{ 1}{4\pi} \nu\Delta\log\rho\)
  instead of the term  \(\frac{ 1}{4\pi} (\frac12-\nu)\Delta\log\rho\)
follows from  the
effective action of \cite {Kivelson,
Read89}. That approach misses the  shift contribution erroneously featuring the wrong sign of the correction.
\end{thebibliography}
\end{document}